\documentclass[aps,a4paper,onecolumn,notitlepage,nobibnotes]{revtex4-1}

\usepackage{helvet}
\usepackage{graphicx}
\usepackage{xspace}
\usepackage{amsmath,amssymb,bm}
\usepackage[usenames,dvipsnames]{xcolor}
\usepackage{MnSymbol}
\usepackage{lineno}
\usepackage{multibib}

\setcitestyle{super}

\newcommand*{\citen}[1]{%
	\begingroup
	\romannumeral-`\x 
	\setcitestyle{numbers}%
	\cite{#1}%
	\endgroup
}

\newif\ifdrafttext

\drafttextfalse 
\ifdrafttext \usepackage[colorlinks,urlcolor=black,citecolor=black,linkcolor=black]{hyperref} \else   \usepackage[hidelinks]{hyperref} \fi

\renewcommand{\figurename}{\textbf{Figure}}
\renewcommand{\tablename}{\textbf{Table}}

\makeatletter
\@addtoreset{figure}{myfigure}
\def\@caption@fignum@sep{~$\vert$~}%
\def\fnum@figure{\textbf{\figurename~\thefigure}}
\makeatother

\makeatletter
\@addtoreset{table}{mytable}
\def\@caption@tablenum@sep{~$\vert$~}%
\def\tnum@table{\textbf{\tablename~\thetable}}
\makeatother

\newcommand{\red}[1]{{\color{black}{#1}}}

\newcommand{\titleofpaper}{Quantum Circuit Refrigerator}
\newcommand{\QCDaffiliation}{QCD Labs, COMP Centre of Excellence, Department of Applied Physics, Aalto University, P.O. Box 13500, FI--00076 Aalto, Finland}

\begin{document}
\title{\titleofpaper}

\author{Kuan Yen Tan}

\author{Matti Partanen}

\author{Russell E. Lake}

\author{Joonas Govenius}

\author{Shumpei Masuda}

\author{Mikko M\"ott\"onen}
\affiliation{\QCDaffiliation}

\date{\today}

\begin{abstract}

Quantum technology promises revolutionizing applications in information processing, communications, sensing, and modelling. However, efficient on-demand cooling of the functional quantum degrees of freedom remains a major challenge in many solid-state implementations, such as superconducting circuits. Here, we demonstrate direct cooling of a superconducting resonator mode using voltage-controllable quantum tunneling of electrons in a nanoscale refrigerator. This result is revealed by a decreased electron temperature at a resonator-coupled probe resistor, even when the electrons in the refrigerator itself are at an elevated temperature. Our conclusions are verified by control experiments and by a good quantitative agreement between a detailed theoretical model and the direct experimental observations in a broad range of operation voltages and phonon bath temperatures. In the future, the introduced refrigerator can be integrated with different quantum electric devices, potentially enhancing their performance. For the superconducting quantum computer, for example, it may provide an efficient way of initializing the quantum bits.

\end{abstract}

\maketitle
\

\section*{Introduction}\label{sec:Intro}

Engineered quantum systems have shown great potential in providing a spectrum of devices superior to the present state of the art in information technological applications.
Since quantum technological devices operate at the level of single energy quanta, they exhibit very low tolerance against external perturbations. Consequently, they need to be extremely well isolated from all sources of dissipation during their quantum coherent operation. These properties typically lead to an elevated operation temperature and long natural initialization times. Thus finding a versatile active refrigerator for quantum devices is of great importance.

One of the greatest challenges of this century is to build a working large-scale quantum computer~\cite{LaddNature2010,Morton2011}. To date, a superconducting quantum computer~\cite{DevoretScience2013} has reached the required gate and measurement accuracy thresholds for fault-tolerant quantum error correction~\cite{Barends2014,Kelly2015} [see also ref.~\citen{Veldhorst2014}]. This device \textcolor{black}{builds} on the decade-long development of circuit quantum electrodynamics~\cite{BlaisPRA2004,WallraffNature2004,DiCarloNature2009,ClarkeNature2008} (cQED), i.e., the study of superconducting quantum
bits, qubits, coupled to on-chip microwave resonators~\cite{KochPRA2007,ManucharyanScience2009,NatafPRL2011}. Although several methods have been demonstrated to initialize superconducting qubits\textcolor{black}{\cite{Valenzuela2006,Reed2010,Mariantoni2011,Riste2012,Geerlings2013,Campagne2013,Bultink2016}} and resonators~\cite{SchliesserNatPhys2008}, \textcolor{black}{they are typically suited only for very specific type of a system and the achieved fidelities fall below} the demanding requirements of efficient fault-tolerant quantum computing. Thus cQED provides an ideal context for the demonstration of a quantum refrigerator.

Electronic microcoolers based on normal-metal--insulator--superconductor (NIS) tunnel junctions~\cite{GiazottoRMP2006, CourtoisJLTP2014} offer opportunities to cool electron systems well below the temperature of the phonon bath even at macroscopic sizes~\cite{LowellAPL2013}. Due to the ideally exponential tunability of the NIS cooling and input powers using an applied bias voltage, these tunnel junctions are attractive candidates for quantum refrigerators. Although single-charge tunneling has previously been demonstrated to emit energy quanta\textcolor{black}{\cite{HofheinzPRL2011,LiuPRL2014,StockklauserPRL2015}} even in applications such as the quantum cascade laser\cite{FaistScience1994}, and \red{artificial-atom masers}\cite{AstafievNature2007,LiuScience2015,Mendes2016}, it has not been experimentally  utilized to directly cool engineered quantum circuits. Even the recently demonstrated autonomous Maxwell's demon has only been used to refrigerate dissipative electron systems~\cite{KoskiPRL2015}.

\red{In this work, we utilize photon-assisted electron tunneling to cool a prototype superconducting quantum circuit, namely a transmission line resonator. The tunneling takes place in NIS tunnel junctions which are coupled to the fundamental resonator mode through voltage (Fig.~\ref{fig1}). We refer to these junctions and their coupling circuitry as the quantum circuit refrigerator (QCR) since it is also integrable with other electric quantum devices.}
The \red{photon-assisted cooling of the resonator} mode is evident from our qualitative observation that a distant probe resistor electrically coupled to the resonator cools down even if the temperature of the QCR,
and of the other heat baths, is elevated (shaded region in Fig.~\ref{fig2}a).
This claim is reinforced by the absence of cooling in a control sample, in which the coupling between the probe and the microwave resonator is suppressed (Fig.~\ref{fig3}, b and c). Furthermore, we obtain a good quantitative agreement between our theoretical model and the experimental results over a broad range of QCR operation voltages and bath temperatures (Figs.~\ref{fig2} and~\ref{fig3}), providing firm evidence of our conclusions. \red{We also verify that the resonator has a well-defined resonance (Fig.~\ref{fig4}).}

\section*{Results}\label{sec:Results}

\subsection*{Experimental samples}\label{subsec:Samples}

\begin{figure}[tbh!] \center
	\includegraphics[width=1\linewidth]{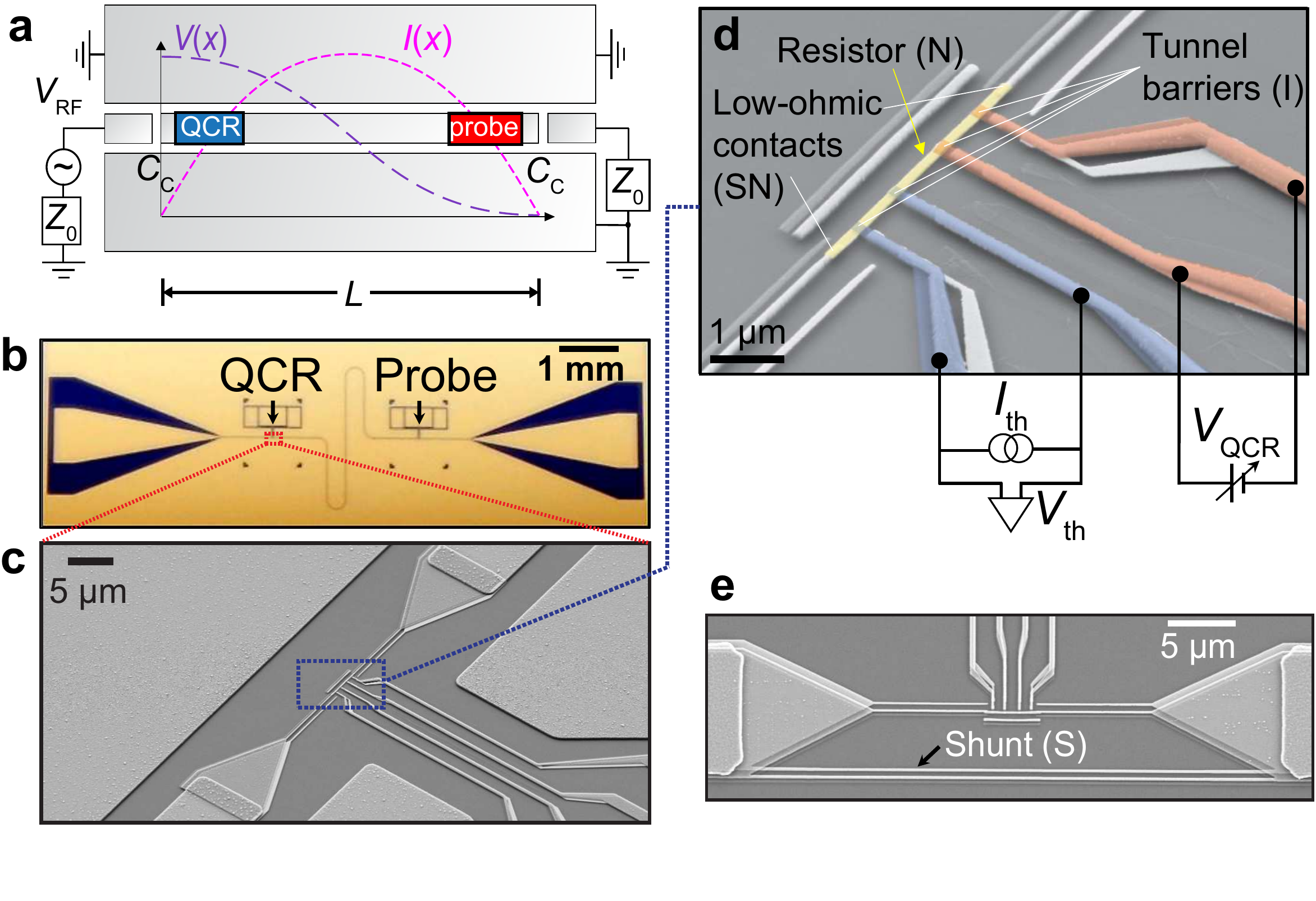}
	\caption{ \label{fig1} \textbf{Experimental sample and measurement scheme.} \textbf{a}, Schematic illustration of the active sample (not to scale) composed of a coplanar-waveguide resonator with an embedded QCR and probe resistor. The voltage, $V(x)$ and current, $I(x)$ profiles of the fundamental mode are shown together with the possibility to apply an external microwave drive, $V_{\mathrm{RF}}$, to the resonator through a coupling capacitor $C_\mathrm{c}$. \textbf{b}, Optical micrograph of an active sample corresponding to \textbf{a}. The QCR and the probe resistor are indicated by the arrows. \textbf{c}, Scanning electron microscope (SEM) image in the vicinity of the QCR. \textbf{d}, Colored SEM image of a QCR with normal-metal (N), insulator (I), and superconductor (S) materials indicated. The refrigerator is operated in voltage bias, $V_\mathrm{QCR}$, while the electron temperature of the normal metal is obtained from the voltage $V_\mathrm{th}$ across a pair of NIS, junctions biased with current $I_\mathrm{th}$.
	\textbf{e}, SEM image of the shunted control sample in the vicinity of the QCR. }
\end{figure}

Figure~\ref{fig1}a--d shows the active sample where a QCR and a probe resistor are embedded near the opposite ends of a superconducting coplanar-waveguide resonator. The refrigerator involves a pair of NIS junctions biased using an operation voltage $V_\mathrm{QCR}$. The probe resistor and the QCR are both equipped with an additional pair of current-biased NIS junctions.  Using a calibration against the bath temperature~(Supplementary Fig.~\ref{figS1}), the observed voltage excursions across these thermometer junctions provide us independent measures of the electron temperatures of the QCR, $T_\mathrm{QCR}$, and of the probe resistor, $T_\mathrm{probe}$. Figure~\ref{fig1}e shows a control sample which has additional superconducting wires in parallel with the QCR and with the probe resistor to decouple these from the electric currents associated with the resonator modes with no other significant effects. Thus the difference between the behaviour of the active sample and that of the control sample can be attributed to microwave photons in the resonator.

\textcolor{black}{The most important device parameters extracted from the experiments are listed in Table~\ref{tab_1}, including  the length $L$ and resonance frequency $f$ of the resonator. The probe resistance $R$ and its distance $x$ from the resonator edge determines the strength of the ohmic coupling between the probe and the resonator (Supplementary Information)}.

\subsection*{Quantum circuit refrigeration}\label{subsec:QCR}

Figure~\ref{fig2}a shows the changes in the electron temperatures of the QCR and of the probe resistor as functions of the QCR operation voltage. Since single electrons cannot tunnel from the QCR into the superconductor unless they overcome the energy gap $2\Delta=430$~$\mu$eV in the superconductor density of states (see Fig.~\ref{fig2}b), the electron temperatures stay essentially unchanged for operation voltages well below $2\Delta/e$, where $e$ is the elementary charge. Slightly below the gap voltage $2\Delta/e$ however, both electron temperatures are significantly decreased. Here, the high-energy electrons at the QCR overcome the superconductor energy gap and tunnel out of the normal metal, thus evaporatively cooling it. Typically, the observed temperature drop at the probe resistor would be simply explained by conduction of heat from it to the QCR. However, this explanation is excluded by our observation that at operation voltages slightly above the gap voltage, the electron temperature in the QCR is well elevated but the probe resistor remains cooled.

Figure~\ref{fig2}b schematically shows different tunneling processes at the QCR with the operation voltage slightly exceeding the superconductor energy gap. Here, the majority of the tunneling events take place near the edges of the gap where the density of states in the superconductor achieves its maximum value. Because the elastic tunneling events, for which the electrons do not exchange energy with the resonator, dominate (Supplementary Information), the normal metal mostly receives electrons above its Fermi level and loses electrons below it. Thus the QCR heats up. However, the electromagnetic mode of the resonator is not affected by elastic tunneling. Instead, the mode cools down by the mechanism of photon-assisted tunneling~\cite{Ingold92,Pekola2010} when it promotes a low-energy electron from the normal metal to the superconductor. On the other hand, the resonator mode heats up when receiving a photon during tunneling of a high-energy electron from the normal-metal to the superconductor. However, this tunneling rate is exponentially suppressed due to the low thermal occupation of high-energy electrons. Thus the resonator mode cools down, resulting in the experimental observations shown in Fig.~\ref{fig2}a: reduced electron temperature at the probe resistor although the QCR temperature is elevated. Note that the other essential heat baths for the probe electrons, composing of phonons in the
solid and quasiparticle excitations in the superconductor, cannot explain the observed cooling since they are heated by the operation of the QCR.

The above-discussed observation is the most striking effect of the resonator acting as a structured electromagnetic environment for the tunneling process. In the Supplementary Information, we discuss in detail how photon-assisted tunneling can be used for on-demand control of
dissipation in the resonator \red{(Supplementary Figs.~\ref{figS2}--\ref{figS4}). Namely, how the operation voltage can be used to exponentially tune the photon lifetime. In contrast to the operation point in Fig.~\ref{fig2}b, a typical operation voltage corresponding to a low effective temperature of this tunable dissipation is well below the gap voltage. Hence the excess heat and quasiparticle generation due to elastic tunneling are greatly suppressed when optimally refrigerating a high-quality quantum device.}

\subsection*{Thermal model}\label{subsec:Model}

To analyze quantitatively the observed temperature changes in Fig.~\ref{fig2}a, we introduce a thermal model shown in Fig.~\ref{fig2}c. We theoretically model the photon-assisted tunneling using the so-called $P(E)$ theory~\cite{Ingold92} for NIS tunnel junctions~\cite{Pekola2010}.
The dominating energy flows into the conduction electrons of the probe resistor are obtained from their coupling to the substrate phonons and to the resonator. The coupling to the resonator arises from ohmic losses \red{in the resistor} due to the electric current associated with the resonator photons~\cite{JonesPRB12}. Similar ohmic \red{losses take place at} the QCR, \red{but this effect is small compared with the desired photon-assisted tunneling if the refrigerator is active}. In addition, we take into account a weak residual heating of the probe resistor due to the power dissipation at the QCR, a constant thermal conductance to an excess bath, and a constant heating \red{of the resonator attributed to} photon leakage from the high-temperature stages of the cryostat. Further details of the thermal model including the employed parameter values are given in Supplementary Information.

\subsection*{Quantum circuit refrigeration explained by the thermal model}\label{subsec:QCRModel}

For a given operation voltage and \red{measured} electron temperature at the QCR, we solve the temperatures of the probe resistor and the resonator from the thermal model (Supplementary Information) such that the different power flows in Fig.~\ref{fig2}c balance each other. The theoretical prediction for the probe temperature is in very good quantitative agreement with our experimental observations as demonstrated in Fig.~\ref{fig2}a. The figure also shows that a theoretical
prediction lacking the contribution from photon-assisted tunneling is in clear \red{disagreement} with the measurements. This is a clear indication that the cooling power of the QCR originates from the direct absorption of resonator photons in the course of electron tunneling.

\subsection*{Effect of bath temperature on the refrigeration}\label{subsec:Temp_dep}

Figure~\ref{fig3}a shows the temperature change of the probe resistor in a broad range of cryostat bath temperatures, $T_0$, and QCR operation voltages. A good quantitative agreement with the experimental data and the thermal model is obtained. For bath temperatures above 200~mK, the QCR operation voltage has a very weak effect on the probe resistor. This loss of probe sensitivity is explained by the quartic temperature dependence of the thermal conductance between the probe electrons and phonons (Supplementary Information). The greater this thermal conductance is, the less sensitive the probe is to the changes of the resonator temperature.

\subsection*{Comparison with a control sample}\label{subsec:Control}

Figure~\ref{fig3}b shows results similar to those in Fig.~\ref{fig3}a but obtained with the control sample, in which the ohmic losses at the QCR and the probe due to the resonator modes are suppressed (Fig.~\ref{fig1}e). Although residual heating is observed at high operation voltages, there is no evidence of refrigeration at the probe. Thus the cooling of the probe in the active sample must arise from the QCR acting on the resonator. This conclusion is also supported by Fig.~\ref{fig3}c, where we show the maximum temperature drop of the probe for the two samples at various bath temperatures. Here, the control sample exhibits no cooling and the theoretical prediction is in very good agreement with the experimental observations.

\subsection*{Microwave response of the resonator}\label{subsec:Microwave}

To verify that the refrigerated resonator has a well-defined mode at the designed  frequency, we introduce rf excitation to one of the input ports of the resonator as described in Fig.~\ref{fig1}a. Although not necessary for the operation of the QCR, these ports are deliberately very weakly coupled to the mode, and hence there is essentially no transmission through the resonator. However, we study the resonance in Fig.~\ref{fig4} by measuring the electron temperatures of the QCR and of the probe resistor as functions of the frequency and amplitude of the excitation. We observe a well-resolved resonance peak centered at $f_0=\omega_0/(2\pi)=9.32$~GHz in agreement with the design parameters (Supplementary Table~\ref{tab_S1}). At the lowest probe powers, the electron temperature and the absorbed power are linearly dependent. Therefore, we can accurately extract the full width at half maximum, $\Delta f_0=70.8$~MHz, using a Lorentzian fit to the electron temperature. The obtained quality factor of the resonance, $Q=f_0/\Delta f_0=132$, indicates that the resonator supports a well-defined mode. Similar experiments on the control sample \red{(Supplementary Fig.~\ref{figS5})} yield a quality factor of 572 indicating that as expected, ohmic losses dominate in the active sample. We attribute the observed losses in the control sample to residual ohmic coupling owing to the finite impedance of the shunt.

\section*{Discussion}\label{Discussion}

\red{The main result of this work is} that single-electron tunneling between a normal metal and a superconductor can be used to refrigerate a microwave resonator on demand.
In the future, the refrigerator will be optimized for minimal ohmic losses.
Such a device can \red{potentially} refrigerate a multitude of \red{high-quality} quantum circuits \red{with conceivable} applications such as precise qubit initialization for large-scale, gate-based quantum computing, \textcolor{black}{quantum-state engineering driven by dissipation~\cite{VerstraeteNatPhys2009}}, and active enhancement of ground-state population in quantum annealing~\cite{AminPRL2008,JohnsonNature2011}. When inactive, the refrigerator is not expected to degrade the quantum coherence as desired. To decrease the base temperature achieved with the device, multiple refrigerators may be cascaded with each other~\cite{CGomezAPL2014} or with distant ohmic reservoirs at very low temperatures~\cite{PartanenNatPhys2016}.

In our experiments, the relatively low quality factor of the resonator is a result of the chosen measurement scheme: rather strong dissipative coupling between the resonator photons and the probe resistor is needed to achieve conveniently measurable temperature changes at the probe resistor when the resonator is being refrigerated. In the future, different probe schemes may be employed where such limitation is absent. These include amplification and analysis of the resonator output signal~\cite{BozyigitNatPhys2011} and measurement of the resonator occupation using a superconducting qubit~\cite{WallraffNature2004,SuriPRA2015}. Furthermore, the ohmic losses due to the QCR may be overcome by minimizing its resistance and by moving it to the end of the resonator where the current profile of the resonator mode ideally vanishes. Importantly, this does not reduce the cooling power of the QCR which utilizes the voltage profile. As discussed in more detail in the Supplementary Information, this type of straightforward improvements of the QCR \red{are expected to allow} for resonator quality factors in the range of 10$^{6}$ when the QCR is inactive~\cite{LindstromJAP2009}. Such level of dissipation \red{is} low enough for the undisturbed operation of quantum devices in their known applications.

\section*{Methods}\label{sec:Methods}

\vspace{10 pt}\noindent{\bf Sample fabrication}

\noindent The samples are fabricated on \red{four-inch} prime-grade intrinsic silicon wafers with 300-nm-thick thermally grown silicon dioxide. The resonator is defined with optical lithography and deposited using an electron beam evaporator, followed by a lift-off process. The evaporated metal film consists three layers from bottom to top: 200~nm of Al, 3~nm of Ti, and 5~nm of Au. Here, gold is used to prevent oxidation and titanium is introduced to avoid the diffusion of gold into the aluminum layer.

The nanostructures are defined by electron beam lithography. Here, we employ a bilayer resist mask consisting of poly(methyl methacrylate) and poly[(methyl methacrylate)-\textit{co}-(methacrylic acid)] to enable three-angle shadow evaporation. The tri-layer nanostructures are deposited in an electron beam evaporator, with in-situ oxidation in between the first layer (Al) and the second layer (Cu) to form the NIS tunnel junctions. The third layer (Al) forms a low-ohmic contact with the second layer which functions as the normal metal in our low-temperature experiments. A lift-off process is performed to remove excess metal. See Table~\ref{tab_1} for the resulting parameter values.

Improvements to the fabrication process towards lowering the amount of electric losses in the resonator are discussed in Supplementary Information.

\vspace{10 pt}\noindent{\bf Measurements}

\noindent For cryogenic electrical measurements, the sample holders are mounted to a cryogen-free dilution refrigerator with a base temperature of 10~mK. The silicon chip supporting the sample is attached with vacuum grease \red{to the sample holder and wedge bonded to the electrical leads of the printed circuit board} using aluminum wires. For each dc line, we employ an individual resistive Thermocoax cable that runs without interruption from the mixing chamber plate of the dilution refrigerator to room temperature.

The NIS thermometers are biased with floating battery-powered current sources and the voltage drops across these junctions are amplified with high-impedance battery-powered voltage preamplifiers before optoisolation and digitalization with an oscilloscope. In the experiments studying different QCR operation voltages, we sweep $\mathit{V}_\mathrm{QCR}$ at a rate of~$\sim$20~$\mu$V/s using an output of an arbitrary function generator that is connected to the cryostat through an optoisolator. The sweep is repeated ten times at each bath temperature.

In the rf measurements, the sinusoidal drive signal is generated by a variable-frequency microwave source and guided to the sample through low-loss coaxial cables which are attenuated at different temperature stages of the cryostat as shown in Supplementary Fig.~\ref{figS1}c.

See Supplementary Information for the details of the NIS thermometry including calibration data.

\vspace{10 pt}\noindent{\bf Modeling}

All numerical computations are carried out using regular desktop computers. See Supplementary Information for a detailed description of the theoretical model used in this work.

\clearpage


\pagebreak
\bibliographystyle{naturemag}

\clearpage
\section*{Acknowledgement}\label{sec:acknowledgement}

The authors thank M. Meschke, J. P. Pekola, and H. Grabert for insightful discussions. This work is supported by the European Research Council under Starting Independent Researcher Grant No. 278117 (SINGLEOUT) and the Academy of Finland through its Centres of Excellence Program (grant nos.\ 251748 and 284621) and grants (nos\ 135794, 272806, 265675, 276528, and 286215), the Emil Altonen Foundation, the Jenny and Antti Wihuri Foundation and the Finnish Cultural Foundation. We also acknowledge the provision of facilities and technical support by Aalto University at Micronova Nanofabrication Centre. The authors declare that they have no competing financial interests.

\section*{Contributions}\label{sec:contributions}
K.Y.T. fabricated the samples, developed and conducted the experiments, and analysed the data. M.P. contributed to the sample fabrication, measurements, and data analysis. R.L. and J.G. contributed to the measurements. S.M. contributed to the theoretical analysis of the system. M.M. provided the initial ideas and suggestions for the experiment, and supervised the work in all respects. All authors discussed both experimental and theoretical results and commented on the manuscript which was written by K.Y.T. and M.M.

\section*{Competing financial interests}
The authors declare no competing financial interests.

\section*{Corresponding authors}
Correspondence should be addressed to M. M\"ott\"onen (mikko.mottonen@aalto.fi) or K. Y. Tan (kuan.tan@aalto.fi).


\clearpage
\section*{Figures and Tables}\label{Figtab}

\clearpage

\begin{figure}[p!] \center
	\includegraphics[width=1\linewidth]{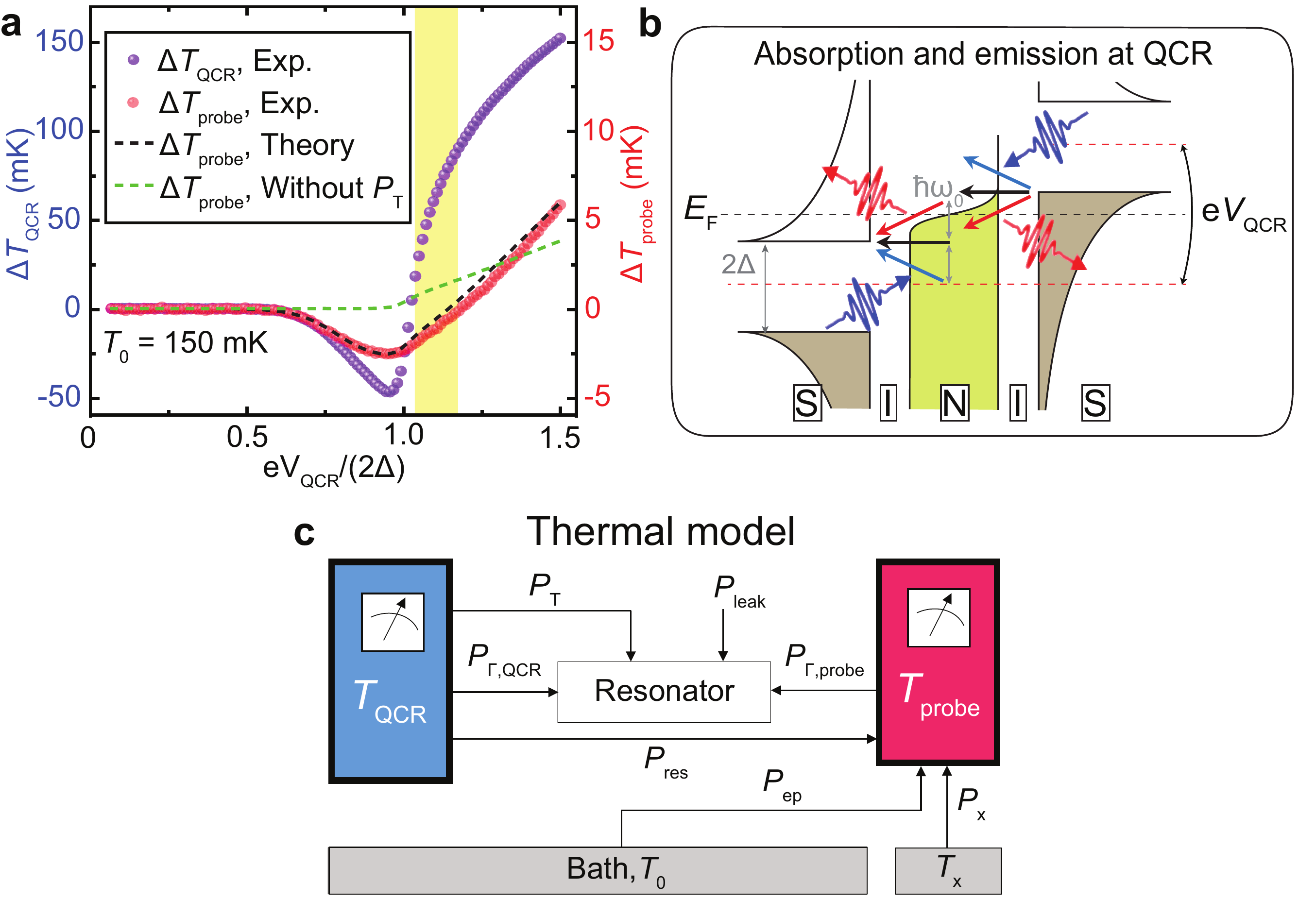}
	\caption{ \label{fig2} \textbf{Quantum circuit refrigeration and thermal model.} \textbf{a}, Changes in the electron temperatures of the QCR, $\Delta T_\mathrm{QCR}$, and of the probe resistor, $\Delta T_\mathrm{probe}$, as functions of the refrigerator operation voltage $V_\mathrm{QCR}$. The black and green dashed lines show the theoretical $\Delta T_\mathrm{probe}$ with and without photon-assisted tunneling, respectively. \textbf{b}, Schematic diagram of photon-assisted single-electron tunneling events in the QCR (blue and red arrows). The wavy arrows denote absorbed (blue) and emitted (red) photons. At the illustrated operation voltage (yellow-shaded region in \textbf{a}) elastic tunnelling (black arrows) dominates, and hence the QCR is heated above the bath temperature  $T_\mathrm{0}$ although the probe exhibits cooling. \textbf{c}, Thermal model of the system illustrating different power flows into the resonator and to the probe. Here, $P_{\mathrm{T}}$ denotes the power  flow into the resonator due to photon-assisted tunnelling; $P_{\mathrm{\Gamma,QCR}}$ and $P_{\mathrm{\Gamma,probe}}$ correspond to ohmic losses in the QCR and probe resistors, respectively; $P_{\mathrm{ep}}$ denotes power transfer owing to coupling of normal-metal electrons to the phonon bath; $P_{\mathrm{res}}$ denotes residual heating power of the probe due to the operation power of the QCR; $P_{\mathrm{leak}}$  accounts for leakage of photons to the resonator from high-temperature stages of the cryostat; and $P_{\mathrm{x}}$  denotes excess power due to a constant thermal conductance $G_{\mathrm{x}}$ to an excess reservoir at temperature $T_{\mathrm{x}}$. Negative power implies the opposite direction of the energy flow with respect to the indicated arrows. See Supplementary Information for further details.}
\end{figure}

\clearpage

\begin{figure}[p!] \center
	\includegraphics[width=1\linewidth]{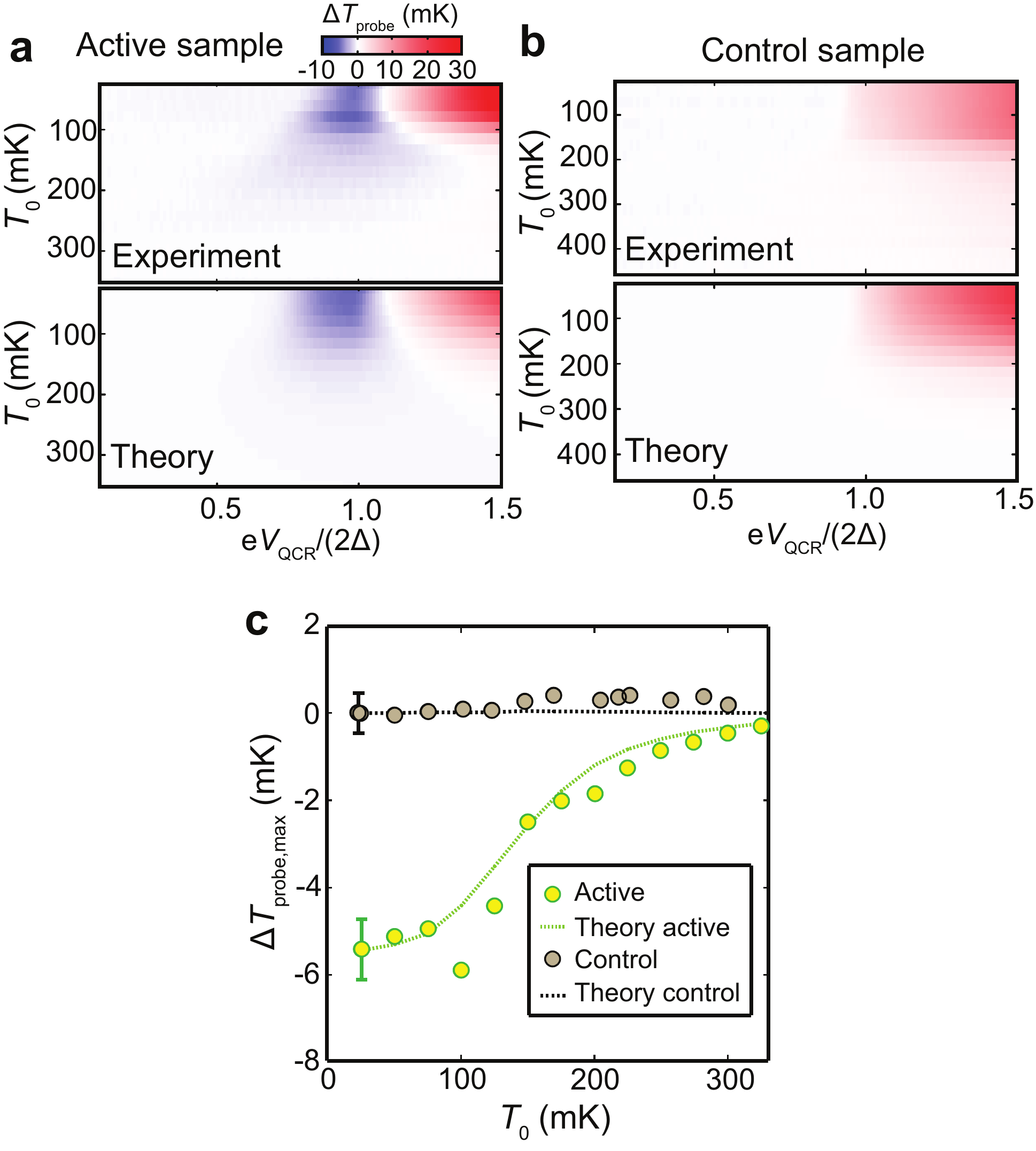}
	\caption{ \label{fig3} \textbf{Comparison between the active sample and the control sample.} \textbf{a} and \textbf{b}, Temperature change of the probe, $\Delta T_\mathrm{probe}$, as a function of the QCR operation voltage and the bath temperature for experimental and simulated data in the case of the active sample (\textbf{a}) and the control sample (\textbf{b}). 
\textbf{c}, Temperature changes of the probe from \textbf{a} and \textbf{b} as functions of the bath temperature at the operation voltage corresponding to the maximum cooling point of the probe and the QCR, respectively. The error bars indicate the \red{maximum} 1$\sigma$ uncertainty for each dataset.}
\end{figure}

\clearpage

\begin{figure}[p!] \center
	\includegraphics[width=0.5\linewidth]{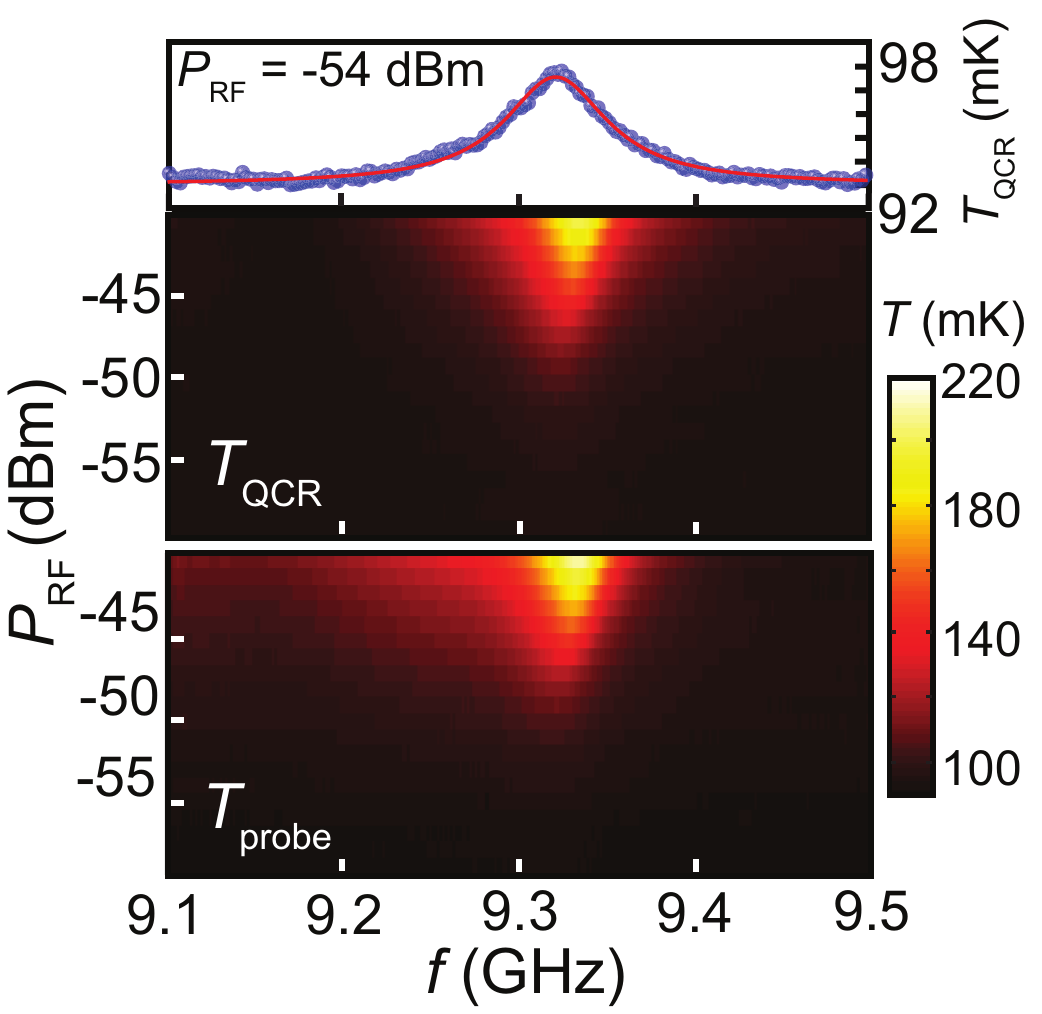}
	\caption{ \label{fig4} \textbf{Observation of the fundamental resonance.} Experimentally observed temperatures of the QCR and of the probe resistor as functions of the frequency and power of the external microwave drive. The measurement scheme is illustrated in Fig.~\ref{fig1}a. The top panel shows a trace of the refrigerator temperature at $-54$-dBm power (markers) together with a Lorentzian fit (solid line). The indicated room-temperature power levels decrease according to Supplementary Fig.~\ref{figS1}c before reaching the sample.}
\end{figure}

\clearpage

\begin{table} [p!]{
		\centering
		\caption[Table caption text]{\textbf{Key device parameters.} These most important device parameters are extracted from the discussed experiments. The full list of parameters used in the thermal model can be found in Supplementary Table~\ref{tab_S1}. \label{tab_1}}
		\begin{tabular}{ l | c | c | c }
			\hline
			Parameter & Symbol & Value & Unit \\
			\hline \hline
			Resonator length& $L$  & 6.833 & mm \\
			Fundamental resonance frequency& $f_0$  & 9.32 & GHz  \\
			Resistance of QCR and probe resistors & $R$  & $46$ & $\Omega$ \\
			Distance of the resistors from resonator edge& $x$  & $100$ & $\mu$m \\
			\hline
		\end{tabular}
		
	}
\end{table}

\clearpage

\setcounter{figure}{0}
\renewcommand*{\thefigure}{S\arabic{figure}}

\setcounter{table}{0}
\renewcommand*{\thetable}{S\arabic{table}}
\setcounter{page}{1}
\renewcommand{\figurename}{\textbf{Supplementary Figure}}
\renewcommand{\tablename}{\textbf{Supplementary Table}}

\setcounter{equation}{0}
\renewcommand*{\theequation}{S\arabic{equation}}
\resetlinenumber[1]
\section*{Supplementary Information}

\vspace{10 pt}\noindent{\bf Temperature calibration in NIS thermometry}\label{Supp:temp_calib}

\noindent The temperatures of the QCR and of the probe are each measured with a pair of current-biased NIS junctions. For elastic single-electron tunneling, the current through an NIS junction with tunneling resistance $R_{\mathrm{T}}$ is given by~\cite{GiazottoRMP2006}
\begin{equation}\label{eq:S1}
I(V,T_\mathrm{e}) =\frac{1}{eR_\mathrm{T}}\int_{-\infty}^{\infty} \mathrm{d}E\,  n_\mathrm{S}(E)[f(E - eV,T_\mathrm{e})-f(E + eV,T_\mathrm{e})]
\end{equation}
where $T_\mathrm{e}\in\{T_\mathrm{QCR},T_\mathrm{probe}\}$ is the electron temperature of the normal-metal resistor, and $V$ the voltage across the NIS junction.
The Fermi--Dirac distribution is given by
\begin{equation}\label{eq:S2}
f(E,T) =\frac{1}{e^{E/(k_\mathrm{B}T)}+1}
\end{equation}
and the quasiparticle density of states in the superconductor can be parametrized by
\begin{equation}\label{eq:S3}
n_\mathrm{S}(E) = \left| \mathrm{Re}\frac{E/\Delta+i\gamma_\mathrm{D}}{\sqrt{(E/\Delta+i\gamma_\mathrm{D})^2-1}} \right|
\end{equation}
The Dynes parameter $\gamma_\mathrm{D}$ \red{accounts, for example, for deep-sub-gap ($|V|\ll\Delta/e$) leakage current.} 
Experimentally, $\gamma_\mathrm{D}$ and $\Delta$ are obtained by fitting equation~\eqref{eq:S1} to the current--voltage characteristics of the NIS junctions used for thermometry.
At sub-gap voltages, equation~\eqref{eq:S1} exhibits a strong dependence on the normal-metal temperature, and hence an NIS junction can be utilized as a secondary thermometer to probe the electron temperature of the normal metal. In our experiments, we apply constant current bias, $I_\mathrm{th,QCR/probe}$, across pairs of NIS junctions and measure the voltage drop across each pair to obtain the signal used to extract the electron temperature.

At high bath temperatures, the electron temperature of the normal metal follows the bath temperature, $T_\mathrm{0}$, giving rise to a faithful conversion function, $g$, of the observed thermometer voltage, $V(T_0)=g^{-1}(T_0)$, into electron temperature, $T_\mathrm{e}=g(V)$. At low temperatures however, the electrons thermally decouple from the phonons leading to a saturation of the electron temperature with decreasing phonon temperature. Before the saturation, the thermometer voltage depends rather linearly on the bath temperature as shown in Supplementary Fig.~\ref{figS1}a. Throughout this paper, we employ such linear conversion from the thermometer voltage to the electron temperature independently for each thermometer.

\vspace{10 pt}\noindent{\bf Thermal Model}\label{Supp:thermal}

\noindent Our thermal model is presented in Fig.~\ref{fig2}c. Several heat transport mechanisms are responsible for the observed temperature of the probe resistor: Firstly, the NIS junctions in the QCR lead to exchange of energy with the resonator due to photon-assisted tunneling, $P_\mathrm{T}$. Secondly, the heat exchange between the normal-metal electrons and the resonator is governed by ohmic losses, $P_{\Gamma,\mathrm{probe}/\mathrm{QCR}}$. Thirdly, the normal-metal electrons are coupled to the phonon bath leading to the power flow $P_\mathrm{ep}$. Fourthly, our model accounts for weak residual heating of the probe due to the power dissipation at the QCR, $P_\mathrm{res}$, and leakage of photons to the resonator from high-temperature stages of the cryostat, $P_\mathrm{leak}$. Finally, we include an excess power $P_\mathrm{x}$ due to a constant thermal conductance $G_\mathrm{x}$ to an excess reservoir at temperature $T_\mathrm{x}$. \red{Supplementary Table~\ref{tab_S1}} shows the values of the parameters used in the model.

In our thermal model (Fig.~\ref{fig2}c), the electron temperature of the probe resistor for a given QCR temperature and operation voltage may be solved from the
power balance equation
\begin{equation}\label{eqn:S4}
P_{\mathrm{\Gamma,probe}}-P_{\mathrm{ep}}-P_{\mathrm{res}}-P_\mathrm{x}=0
\end{equation}
The power flowing from the probe electrons to the resonator photons due to ohmic losses can be expressed as~\cite{JonesPRB12}
\begin{equation}\label{eqn:S5}
P_{\mathrm{\Gamma,probe}}=\Gamma^{\mathrm{probe}}_{0\rightarrow 1} p_0\hbar\omega_0 - \Gamma^{\mathrm{probe}}_{1\rightarrow 0} p_1\hbar\omega_0
\end{equation}
where $\Gamma^{\mathrm{probe}}_{0\rightarrow 1}$ and $\Gamma^{\mathrm{probe}}_{1\rightarrow 0}$ are the excitation and relaxation rates of the resonator photons due to the probe resistor, respectively, and $p_0= 1-p_1$ is the probability of the resonator to be in its quantum-mechanical ground state. For simplicity, we consider here only the two lowest-energy states of the resonator. In the steady state achieved in our experiments, we have
\begin{equation}\label{eqn:S6}
\dot{p}_0=0=-\Gamma^+p_0+\Gamma^-p_1 \quad \Rightarrow \quad p_0=\frac{\Gamma^-}{\Gamma^-+\Gamma^+},\quad p_1=\frac{\Gamma^+}{\Gamma^-+\Gamma^+}
\end{equation}
where the total excitation and relaxation rates of the resonator mode are given by
\begin{equation}\label{eqn:S7}
\begin{split}
\Gamma^+ & = \Gamma^{\mathrm{QCR}}_{0\rightarrow 1} + \Gamma^{\mathrm{probe}}_{0\rightarrow 1} + \Gamma^{\mathrm{T}}_{0\rightarrow 1} + \Gamma_{\mathrm{leak}}\\
\Gamma^- & = \Gamma^{\mathrm{QCR}}_{1\rightarrow 0} + \Gamma^{\mathrm{probe}}_{1\rightarrow 0} + \Gamma^{\mathrm{T}}_{1\rightarrow 0} 
\end{split}
\end{equation}
respectively. Here, the rate $\Gamma_{\mathrm{leak}}$ determines the leakage power to the resonator $P_\mathrm{leak}=\hbar\omega_0 \Gamma_{\mathrm{leak}} p_0$ and the rate $\Gamma^\mathrm{T}$ arises from the photon-assisted tunneling at the QCR as described in the next section. The rates arising from the ohmic losses are given by~\cite{JonesPRB12}
\begin{equation}\label{eqn:S8}
\Gamma^{\mathrm{QCR/probe}}_{0\rightarrow 1}=\frac{\gamma}{e^{\hbar\omega_0/(k_\mathrm{B}T_\mathrm{QCR/probe})}-1},\quad \Gamma^{\mathrm{QCR/probe}}_{1\rightarrow 0}=\frac{\gamma}{1-e^{-\hbar\omega_0/(k_\mathrm{B}T_\mathrm{QCR/probe})}}
\end{equation}
where we employ the same base rate $\gamma=2\omega_0R\sin^2(\pi x/L)/(\pi Z_0)$ for the QCR and the probe due to the symmetry in their resistances $R$ and distances from the edge of the resonator $x$. The characteristic impedance $Z_0=\sqrt{L_\mathrm{l}/C_\mathrm{l}}$ of the resonator of length $L$ is given by the inductance and capacitance per unit length $L_\mathrm{l}$ and $C_\mathrm{l}$, respectively.

Equation~\ref{eqn:S4} also includes the term $P_{\mathrm{ep}}$ that arises from the coupling between the normal-metal electrons and the substrate phonons. This heat flow is given by~\cite{GiazottoRMP2006}
\begin{equation}\label{eqn:S9}
P_{\mathrm{ep}}= \Sigma_{\mathrm{Cu}}\Omega_{\mathrm{probe}}(T_\mathrm{0}^5-T_\mathrm{probe}^5)
\end{equation}
where $\Omega_{\mathrm{probe}}$ is the volume of the probe resistor and $\Sigma_{\mathrm{Cu}}= 2\times10^9$~W~K$^{-5}$~m$^{-3}$ is the known electron--phonon coupling constant of copper~\cite{GiazottoRMP2006}.

In the control sample, we observed weak heating of the probe due to the power dissipation at the QCR. This heating is approximately linearly dependent on the bath temperature, and hence we include a residual heating power to our thermal model in the form
\begin{equation}\label{eqn:S10}
P_{\mathrm{res}} = \alpha(\beta-T_\mathrm{0})I_\mathrm{QCR}V_\mathrm{QCR}
\end{equation}
We fix the values of the parameters $\alpha$ and $\beta$ (see \red{Supplementary Table~\ref{tab_S1}}) using the measurement data of the control sample and use the same values also in the case of the active sample. Importantly, the contribution of this residual heating is typically much weaker than that of the other heat conduction mechanisms. The microscopic origin of the residual heating remains to be studied further but the existence of such very weak channel is not surprising.

In equation.~\eqref{eqn:S4}, we choose the excess power to assume the form
\begin{equation}\label{eqn:S11}
P_\mathrm{x}=G_\mathrm{x}(T_\mathrm{x}-T_\mathrm{probe})
\end{equation}
where, for simplicity, $G_\mathrm{x}$ is a constant thermal conductance and $T_\mathrm{x}$ is the temperature of the excess bath. We assume that the dominating thermal coupling of the excess bath to the phonon bath is through electron--phonon coupling and that the excess bath is so large that its temperature is essentially independent of the temperature of the probe. However, we assume a constant heating power, $P_\mathrm{x}^\mathrm{con}$, at the excess bath which leads to a finite saturation temperature, $T_\mathrm{x}^\mathrm{sat}$, even at zero phonon bath temperature. Equating $P_\mathrm{x}^\mathrm{con}$ with the power due to the electron--phonon coupling (see equation~\eqref{eqn:S9}) yields $T_\mathrm{x}=[(T_\mathrm{x}^\mathrm{sat})^5+T_0^5]^{1/5}$.

We adjust $G_\mathrm{x}$ and $T_\mathrm{x}^\mathrm{sat}$ to match $T_\mathrm{probe}$ predicted by the thermal model to that measured at bath temperatures $T_0=25$~mK and 50~mK without operating the QCR ($V_\mathrm{QCR}= 0$). Supplementary Fig.~\ref{figS1}b shows the measured probe temperature in this case together with the prediction of the thermal model. Although the model is fitted to the measured data only at the two lowest bath temperatures, very good agreement with the experimental results and the theoretical prediction is achieved in the whole temperature regime, in which the linear temperature calibration is valid.

As described below, there are no free parameters in the resonator excitation and relaxation rates that give rise to the power
\begin{equation}\label{eqn:S12}
P_\mathrm{T}=\hbar\omega_0(\Gamma^\mathrm{T}_{0\rightarrow 1}p_0-\Gamma^\mathrm{T}_{1\rightarrow 0}p_1)
\end{equation}
arising from the photon-assisted tunneling. Thus $\Gamma_\mathrm{leak}$ is the only parameter we adjust to fit the thermal model to the temperature drops observed in Figs.~\ref{fig2} and~\ref{fig3} at the probe due to the QCR. Since we adjust the value of $\Gamma_\mathrm{leak}$ to obtain a good match at the lowest bath temperature $T_0=25$~mK, the results of the thermal model at higher temperatures such as those at $T_0=150$~mK in Fig.~\ref{fig2}a may be considered as a theoretical prediction.

\vspace{10 pt}\noindent{\bf Photon-assisted single-electron tunneling}\label{Supp:PE}

\noindent In our case, the single-electron tunneling through the NIS junctions can be described by means of the Fermi golden rule taking into account the voltage fluctuations arising from the electromagnetic environment of the junction. In this $P(E)$ theory~\cite{Ingold92}, the forward tunneling rate, i.e., the rate for an electron to tunnel from the normal metal to the superconductor, is given by~\cite{Ingold92,Pekola2010}
\begin{equation}\label{eq13}
\vec{\Gamma}(V_\mathrm{QCR})  =\frac{1}{e^2R_\mathrm{T}}\int_{-\infty}^{\infty} \int_{-\infty}^{\infty}\mathrm{d}E\,\mathrm{d}E'\, n_\mathrm{S}(E')f_\mathrm{N}(E - eV_{\mathrm{QCR}})[1-f_\mathrm{S}(E')]P(E-E')
\end{equation}
where the Fermi distribution functions $f_\mathrm{N}(E)=f(E,T_\mathrm{QCR})$ and $f_\mathrm{S}(E)=f(E,T_0)$ are given by equation~\eqref{eq:S2} and $P(E)$ is the probability density function for the environment to absorb $E$ amount of energy. For simplicity, we have assumed above that the quasiparticle excitations in the superconductor are well thermalized with the phonon bath.

In the zero-temperature limit for the fundamental mode of the resonator acting as the environment, we have for an NIS tunnel junction~\cite{Ingold92}
\begin{equation}\label{eqn:S14}
P(E) \approx e^{-\rho} \sum_{k=0}^{\infty}\frac{\rho^k}{k!}\delta(E-k\hbar\omega_0)=\sum_{k=0}^{\infty}q_k\delta(E-k\hbar\omega_0),
\end{equation}
where $\rho$ is an environmental parameter
\begin{equation}\label{eqn:S15}
\rho=\frac{\pi}{CR_\mathrm{K}\omega_0}
\end{equation}
that depends on the effective capacitance, $C=|L/2-x|C_\mathrm{l}/2\approx LC_\mathrm{l}/2$ for $x\ll1$, of the $LC$ oscillator which is used to model the fundamental mode. Here, $R_\mathrm{K}=25.8$~k$\Omega$ is the von Klitzing constant. The coefficient $q_k$ equals the probability of emitting $k$ quanta of energy to the resonator in the course of single-electron tunneling. Since $\rho=4.7\times 10^{-3}\ll 1$ \red{with our parameters} (see \red{Supplementary Table~\ref{tab_S1}}), the elastic tunneling events, $k=0$, for which no heat exchange with the resonator takes place, clearly dominate in the probabilities.

In our case of finite temperature, the probability for the tunneling electron to absorb a quantum of energy from the resonator, $q_{-1}$, is related to the emission probability by the detailed-balance condition $q_{-1}=q_{1}\mathrm{exp}[-\hbar\omega_0 / (k_\mathrm{B}T_\mathrm{r})]$, where $T_\mathrm{r}$ is the temperature of the resonator. For simplicity, we consider only zero- and single-photon events, justified by $\rho \ll 1$.

In the following, we consistently assume the low-temperature limit for the resonator, which is required by the fact that we take only the zero- and single- photon states of the lowest resonator mode into account. Thus we obtain the approximate probabilities
\begin{equation}\label{eqn:S16}
q_0=\frac{1}{1+\rho},\quad q_1=\frac{\rho}{1+\rho}\times\frac{1}{1+e^{-\hbar\omega_0/(k_\mathrm{B}T_\mathrm{r})}},\quad q_{-1}=\frac{\rho}{1+\rho}\times\frac{e^{-\hbar\omega_0/(k_\mathrm{B}T_\mathrm{r})}}{1+e^{-\hbar\omega_0/(k_\mathrm{B}T_\mathrm{r})}}
\end{equation}
and
\begin{equation}\label{eqn:S17}
P(E)=\sum_{k=-1}^{1}q_k\delta(E-k\hbar\omega_0)
\end{equation}

Using equations~\eqref{eq13},~\eqref{eqn:S16}, and~\eqref{eqn:S17}, we obtain the tunneling rates for the electrons in the forward direction, $\overset{\rightarrow}{\Gamma}$. The backward rate $\overset{\leftarrow}{\Gamma}$ can be obtained in a similar fashion~\cite{Ingold92}. The rates can further be expressed as sums of contributions from the different processes: emission ($\overset{\rightarrow}{\Gamma}_{1}$ and $\overset{\leftarrow}{\Gamma}_{1}$), absorption ($\overset{\rightarrow}{\Gamma}_{-1}$ and $\overset{\leftarrow}{\Gamma}_{-1}$), and elastic tunneling ($\overset{\rightarrow}{\Gamma}_{0}$ and $\overset{\leftarrow}{\Gamma}_{0}$). These electron tunneling rates are distinct from the photon-assisted resonator excitation and relaxation rates in equation~\eqref{eqn:S7} which can be expressed as
\begin{equation}\label{eqn:S18}
\begin{split}
{\Gamma}{^\mathrm{T}_\mathrm{0\rightarrow 1}}&
=\overset{\rightarrow}{\Gamma}{^\mathrm{T}_\mathrm{0\rightarrow 1}}
+\overset{\leftarrow}{\Gamma}{^\mathrm{T}_\mathrm{0\rightarrow 1}}\\
{\Gamma}{^\mathrm{T}_\mathrm{1\rightarrow 0}}&
=\overset{\rightarrow}{\Gamma}{^\mathrm{T}_\mathrm{1\rightarrow 0}}
+\overset{\leftarrow}{\Gamma}{^\mathrm{T}_\mathrm{1\rightarrow 0}}
\end{split}
\end{equation}
where $\overset{\rightarrow}{\Gamma}{^\mathrm{T}_\mathrm{0\rightarrow 1}}=\overset{\rightarrow}{\Gamma}{_\mathrm{1}}/p_0$, $\overset{\leftarrow}{\Gamma}{^\mathrm{T}_\mathrm{0\rightarrow 1}}=\overset{\leftarrow}{\Gamma}{_\mathrm{1}}/p_0$, $\overset{\rightarrow}{\Gamma}{^\mathrm{T}_\mathrm{1\rightarrow 0}}=\overset{\rightarrow}{\Gamma}{_\mathrm{-1}}/p_1$, and $\overset{\leftarrow}{\Gamma}{^\mathrm{T}_\mathrm{1\rightarrow 0}}=\overset{\leftarrow}{\Gamma}{_\mathrm{-1}}/p_1$ since the average number of electrons tunneled in a given process is equal to the average number of photons exchanged in this process. Here, the direct dependence of the \red{resonator excitation and relaxation} rates on the temperature of the resonator
is canceled by the temperature dependence of the resonator populations (see equation~\eqref{eqn:S6}).
Thus the resonator experiences the QCR as a voltage-tunable environment. Using the above results, the forward resonator rates assume the forms
\begin{equation}\label{eqn:S19}
\begin{split}
\overset{\rightarrow}{\Gamma}{^\mathrm{T}_\mathrm{0\rightarrow 1}}&
=\frac{\rho}{1+\rho}\frac{1}{e^2R_\mathrm{T}}\int_{-\infty}^{\infty}\mathrm{d}E\, f_\mathrm{N}(E-eV)n_\mathrm{S}(E-\hbar\omega_0)[1-f_\mathrm{S}(E-\hbar\omega_0)]\\
\overset{\rightarrow}{\Gamma}{^\mathrm{T}_\mathrm{1\rightarrow 0}}&
=\frac{\rho}{1+\rho}\frac{1}{e^2R_\mathrm{T}}\int_{-\infty}^{\infty}\mathrm{d}E\, f_\mathrm{N}(E-eV)n_\mathrm{S}(E+\hbar\omega_0)[1-f_\mathrm{S}(E+\hbar\omega_0)]
\end{split}
\end{equation}
Thus the photon-assisted resonator relaxation and excitation rates (equation~\eqref{eqn:S18}) can be theoretically predicted without any free parameters. Consequently, these rates affect the power flowing into the probe resistor through equations~\eqref{eqn:S5}--\eqref{eqn:S7}.

\vspace{10 pt}\noindent{\bf Minimizing undesired losses due to the QCR }\label{Supp:QF}

Internal quality factors of a bare superconducting coplanar-waveguide resonators, $Q_{\mathrm{int,bare}}$, of the order $\sim10^6$ have been demonstrated in the single-photon regime~\red{\cite{LindstromJAP2009,VissersAPL2010,Bruno2015}}.  Such state-of-the-art values may be obtained with sophisticated fabrication techniques employing proper choices of materials such as TiN on a high-purity silicon substrate~\cite{VissersAPL2010,OhyaSST2014,JaimIEEE2015}. In this section, we discuss the sources of dissipation added by the introduction of the QCR into the resonator and give a sample design which is optimized for low losses although not hindering the desired operation characteristics of the QCR. Using realistically achievable parameters, \red{our analysis indicates} that the optimized design \red{allows us to make the additional losses due to the QCR small} compared with an internal quality factor of $10^6$. Importantly, the optimized design is also compatible with the fabrication techniques of the low-loss resonators and other superconducting quantum devices, and hence the QCR holds great potential in introducing temporally controlled dissipation without degrading the coherence properties when inactive.

In this paper, we measure the resonator temperature using a probe resistor that couples through ohmic losses to the resonator. However, such dissipative measurement technique is not necessary in the future. For example, if the QCR is used to cool a high-quality resonator, the photon occupation numbers may be measured using a dispersively coupled superconducting qubit~\cite{SuriPRA2015}. Thus we consider below a case, in which there is no probe resistor in the system. In this case, we differentiate three possible sources of dissipation: ohmic losses at the QCR, losses due to the smearing of the superconductor density of states, and losses at the metal insulator interfaces. We discuss each of these below. In addition, we investigate \red{in the next section} the losses owing to the photon-assisted tunneling giving rise to the operation of the QCR.
Classically, the normal-metal resistor, $R$, of the QCR introduces dissipation in the resonator mode due to the electric current, $I(x)$, carried by the excitations of the mode and the Ohms law $P_{\mathrm{res}}=RI^2(x)$. Thus it is natural that the ohmic losses can be greatly reduced by reducing the resistance value and moving the resistor close to the end of the resonator where the current profile of the mode linearly vanishes. Employing the quantum-mechanical treatment used in equation~\eqref{eqn:S8}, the internal quality factor due to this loss mechanism only assumes the form~\cite{JonesPRB12}

\begin{equation}\
\label{eqn:S20}
Q_{\mathrm{int,\red{ohm}}}\approx\frac{\omega_0}{\gamma}
\end{equation}

\noindent where $\gamma=2\omega_0R\sin^2(\pi x/L)/(\pi Z_0)$ is the resonator internal dissipation rate due to ohmic losses. For the optimized sample design shown in Supplementary Fig.~\ref{figS2}a, the resistor is at the very end of the resonator, and hence corrections to the mode current profile from the total junction capacitance, $C_{\mathrm{J}}^{\mathrm{tot}}$, are significant. Thus \red{we} estimate the effective distance from the resonator end to be given by $x = C_{\mathrm{J}}^{\mathrm{tot}}/C_{\mathrm{l}} = 3.4~\mu$m. Together with the resistance $R = 0.3~\Omega$ of the copper block this implies $Q_{\mathrm{int,\red{ohm}}} = 1.3\times10^8$. Another way to arrive at an equal $Q_{\mathrm{int,\red{ohm}}}$ is the following: (i) treat the resistor and the junction capacitors as a lumped-element termination impedance, $Z$, for the resonator, (ii) calculate the current through the impedance, $I_\mathrm{z}$, using the impedance $Z$ and the undisturbed voltage of the resonator mode, and (iii) obtain the dissipated power from Ohm\textquotesingle s law, $RI_\mathrm{z}^2$. Hence these losses have a negligible effect on the total internal quality factor of the resonator assuming that $Q_{\mathrm{int,\red{bare}}}=10^6$ .

The Dynes density of states for the tunnel junctions of the QCR may also contribute to the internal loss of the resonator. For a single NIS junction, we estimate this loss as
\begin{equation}\
\label{eqn:S21}
Q_{\mathrm{int,Dynes}}=2\pi\times\frac{n\hbar\omega_0^2}{P_{\mathrm{Dynes}}/f_0}=\frac{n\hbar\omega_0^2}{P_{\mathrm{Dynes}}}
\end{equation}

\noindent where $P_{\mathrm{Dynes}}=(\langle\hat{V}_{\mathrm{res}}\rangle^2-\mathrm{z.p.f})/R_{\mathrm{Dynes}}$ is the photon power dissipation due to the subgap resistance $R_{\mathrm{Dynes}}=R_\mathrm{T}/\gamma_\mathrm{D}$, $n$ is the average photon number, and z.p.f. denotes the contribution arising from the zero-point voltage fluctuations. The voltage operator of the resonator fundamental mode  $\hat{V}_{\mathrm{res}}$  is given by

\begin{equation}\
\label{eqn:S22}
\hat{V}_{\mathrm{res}}=\sqrt{\frac{\hbar\omega_0}{LC_\mathrm{l}}}(\hat{a}+\hat{a}^\dagger)\mathrm{cos}(\frac{\pi x}{L})
\end{equation}

\noindent where $\hat{a}$ and $\hat{a}^\dagger$ represent the bosonic annihilation and creation operators \red{of the mode}, respectively. Thus we may express equation~\eqref{eqn:S21} as
	
\begin{equation}\
\label{eqn:S23}
 Q_{\mathrm{int,\red{Dynes}}}=\frac{R_\mathrm{T}LC_\mathrm{l}\omega_0}{2\gamma_\mathrm{D}}\mathrm{cos}^{-2} (\frac{\pi x}{L})\approx\frac{\pi R_\mathrm{T}}{2\gamma_\mathrm{D} Z_0}
\end{equation}

\noindent where the cosine term is approximated to be unity for $x \ll 1$ . Even for the sample realized in this paper, equation~\eqref{eqn:S23} yields a Dynes quality factor well above $10^6$. Thus these losses are negligible for the optimized sample, the parameters of which yield a Dynes quality factor above $10^{10}$ because of the larger tunneling resistance and smaller Dynes parameter. A typical reason for the smearing of the density of states is photon-assisted tunneling arising from noise coupled through the dc leads of the NIS junctions, which can be suppressed by introducing shunt capacitors to ground~\cite{Pekola2010}. This additional photon-assisted tunneling may be treated independent of that due to the resonator photons since the voltage fluctuations related to these two mechanisms are uncorrelated.

In addition to the above-discussed loss mechanisms, in principle, there may be additional dissipation arising from the normal-metal--insulator interfaces which have not yet been thoroughly investigated in the context of cQED. Typically, such losses are attributed to quantum fluctuators coupling to the voltage drop across the interface. Since we expect such loss mechanisms to be very weak, we have utilized in the optimized design capacitive coupling of the normal-metal to the resonator. Instead of the parallel-plate design, a finger capacitor may be used as well.  Furthermore, the normal metal may as well be galvanically connected to the superconducting resonator \red{implying essentially} no voltage drop at the arising normal-metal--superconductor interface due to the large series impedance of the NIS tunnel junction, and hence no losses arise from possible fluctuators here. Although the capacitor at the NIS junction cannot be removed, the junction can be fabricated purely from the same type of aluminum as is used in the Josephson junctions of typical cQED architectures by employing the inverse proximity effect as described in \red{ref.~\onlinecite{KoskiAPL2011}.} Thus it is possible to distinguish the any unwanted dissipation arising from normal-metal--insulator interfaces.

In summary, the optimized sample design shown in Supplementary Fig.~\ref{figS2} is expected to add insignificant amount of undesired dissipation to the resonator mode. Since a resonator is a central component in cQED and many the state-of-the-art qubits can be described as slightly anharmonic oscillators, these estimates suggests that the QCR can be used in the future to directly cool a multitude of quantum technological components.

\vspace{10 pt}\noindent{\bf Temporal control of dissipation using the QCR}\label{Supp:Temporal}

The QCR appears to the coupled quantum device as a dissipative environment, the temperature and the coupling strength of which can be temporally controlled using the operation voltage. Supplementary Fig.~\ref{figS3}a shows the excitation ($\Gamma_{0\rightarrow1}^{\mathrm{T}}$) and relaxation ($\Gamma_{1\rightarrow0}^{\mathrm{T}}$) rates of the resonator mode due to the photon-assisted tunneling at the QCR. Even for the parameters of the measured sample, this photon-assisted tunneling rate gives rise to very weak dissipation at vanishing operation voltage in comparison to a bare internal quality factor of $10^6$ \red{(see Supplementary Fig.~\ref{figS3}b)}. However, the rates increase exponentially with the operation voltage providing the possibility of a fast reset of the resonator mode when desired. We find in Supplementary Fig.~\ref{figS3}a an optimal operation voltage with respect to the temperature corresponding to the photon-assisted tunneling. This optimal voltage depends on the Dynes parameter and the electron temperature of the normal metal. After one reaches the desired temperature of the refrigerated quantum device, it is beneficial to quickly ramp down the QCR operation voltage to zero.

\vspace{10 pt}\noindent{\bf Temperature of the resonator during refrigeration}\label{Supp:Tres}

\noindent At high operation voltages, the QCR is not cooling the resonator mode but substantially heating it. In this operation regime, we expect to have \red{considerable} multi-photon occupation in the mode, which is not accurately captured by the two-state approximation employed above. Thus for an improved accuracy in the estimation of the resonator temperature and of the average photon number, we utilize an upgraded thermal model which includes also the multi-photon states.

In this upgraded model, equations~\eqref{eqn:S5}--\eqref{eqn:S8}, \eqref{eqn:S12}, and \eqref{eqn:S15} are replaced by the corresponding equations containing occupation probabilities of all photon number states $p_\mathrm{k}$. Consequently, ohmic losses induce transitions between the adjacent number states, and although rare, photon-assisted tunneling may induce multi-photon absorption or emission. By invoking the typical assumption used in $P(E)$ theory that the resonator is in a thermal state, we may express the power flows into and out of the resonator using its temperature, or equivalently the average photon number

\begin{equation}\
\label{eqn:S24}
n=\frac{1}{\mathrm{exp}[\hbar\omega_0/(k_\mathrm{B}T_{\mathrm{res}})]-1}
\end{equation}

\noindent In the simulation, the average photon number changes according to

\begin{equation}\
\label{eqn:S24}
 \hbar\omega_0\delta_\mathrm{t}n=P_\mathrm{T}+P_\mathrm{\Gamma,QCR}+P_\mathrm{\Gamma,probe}+P_\mathrm{leak}
\end{equation}

\noindent until a stationary state is achieved. In the stationary state, the power flows (see Fig.~\ref{fig2}c) balance each other.

Supplementary Fig.~\ref{figS4} shows the results of the upgraded thermal model: temperature and average photon number of the fundamental mode of the resonator as a function of the QCR operation voltage. At maximum cooling, the average photon number is reduced down to \red{$n$ = 0.3} using parameters corresponding to the measured sample. This number is well above the state of the art in cQED. However, the aim of this work is not to show record-low photon numbers but to demonstrate the principle of quantum circuit refrigeration. In fact, a relatively high initial photon number is beneficial in our experiments since is renders the temperature drop of the probe resistor observable when the QCR is operated. \textcolor{black}{Nevertheless, the QCR is theoretically expected operate also in a state-of-the-art setup with very small photon leakage rates, possibly allowing for extremely low temperatures after refrigeration.}

Although Supplementary Fig.~\ref{figS4} shows that the two-level approximation for the resonator mode is compromised at certain operation regimes of the QCR, \red{this simple model} captures the essential physics well and provides a quantitative match with the experimental results with as many fitting parameters as the upgraded model. Furthermore, we verified that the upgraded many-state model also yields a good quantitative match with the experiments leading to the same conclusions as the original thermal model. Thus for simplicity, we choose to work within the two-state model in the main body of this work.

\begin{figure}[p!] \center
	\includegraphics[width=0.8\linewidth]{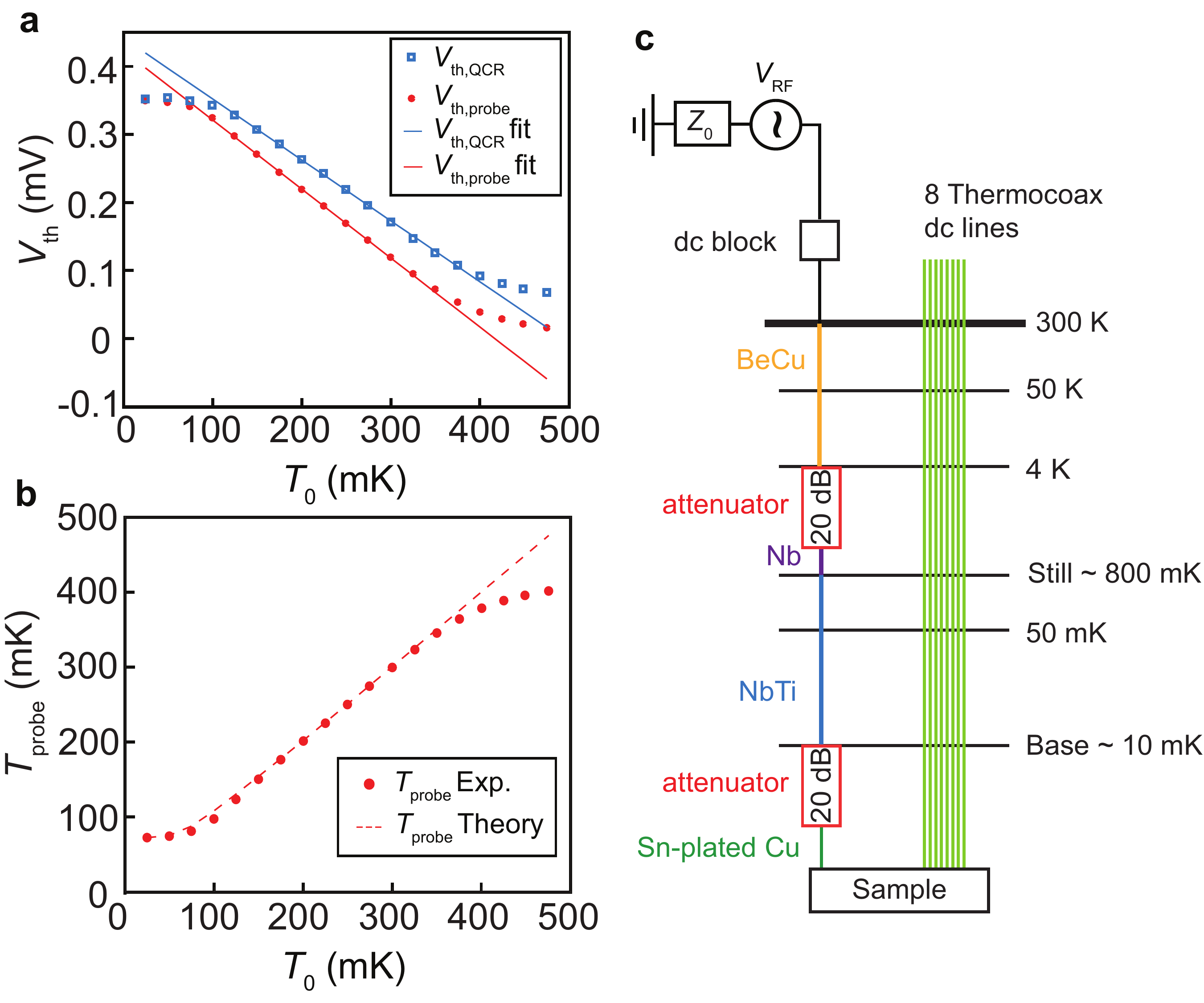}
	\caption{ \label{figS1} \textbf{Thermometer calibration and experimental wiring.} \textbf{a}, Thermometer voltages of the QCR and of the probe as functions of the phonon bath temperature. The solid lines are linear fits to the experimental data (dots) and they are used to convert the measured thermometer voltages into the electron temperatures at the QCR and at the probe. \textbf{b}, Measured electron temperature of the probe using the calibration from \textbf{a} as a function of the bath temperature. The dashed line shows the electron temperature extracted from the thermal model of Fig.~\ref{fig2}c. The deviation of the theoretical prediction from the measurement data at high bath temperatures is due to the failure of the linear thermometer calibration also visible in \textbf{a}. In both panels, the QCR operation voltage is set to zero. \textbf{c}, Wiring scheme for the measurements. For the rf signal, 20-dB attenuators are attached at different temperature stages of the cryostat for improved thermalization. Below 4-K temperature, superconducting coaxial cables are used. Resistive Thermocoax cables are employed for the dc lines. }
\end{figure}
\clearpage

\begin{figure}[p!] \center

	\includegraphics[width=1\linewidth]{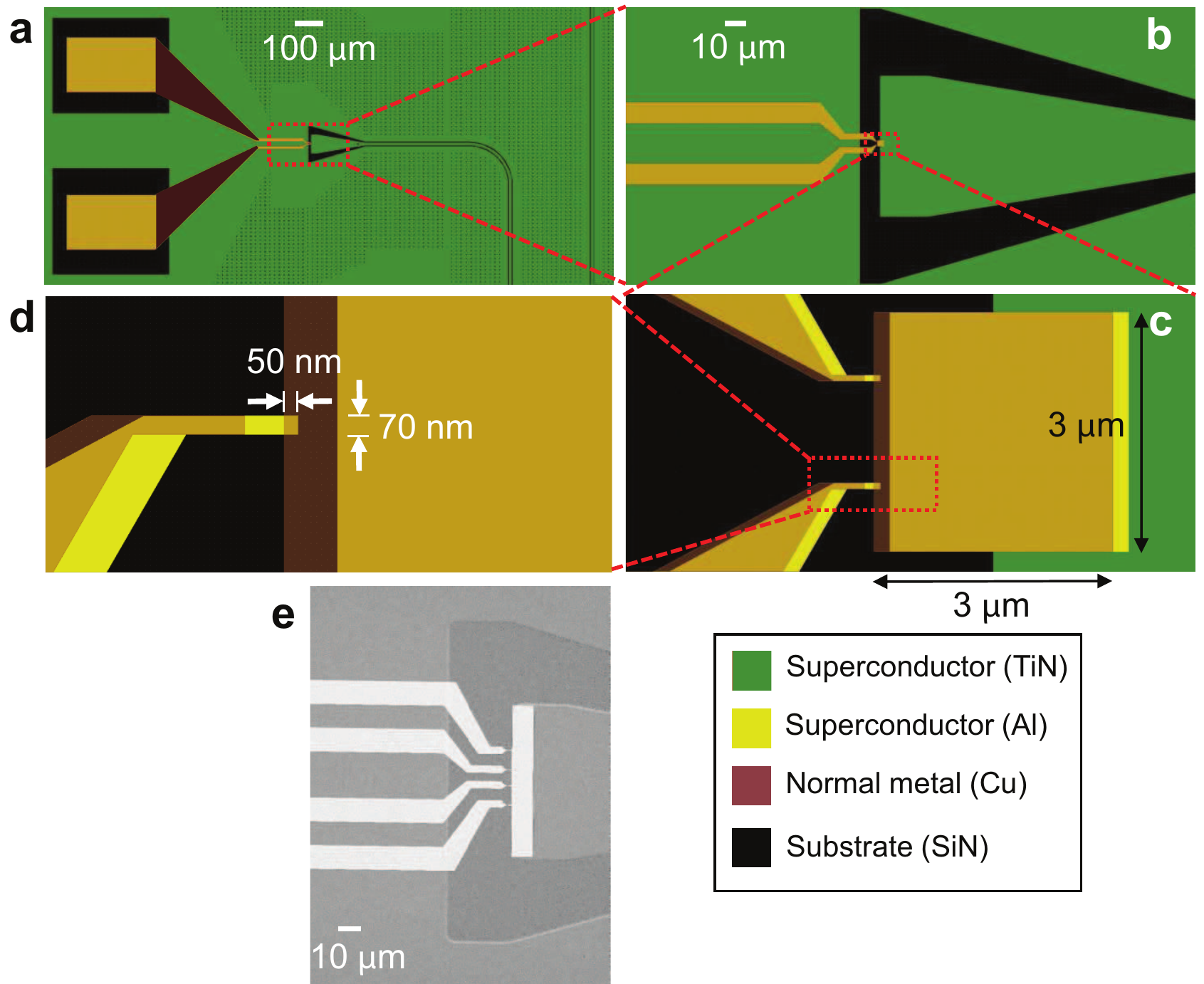}
	\caption{ \label{figS2} \textbf{Optimized sample design.} \textbf{a}, Quantum circuit refrigerator with two dc bias leads and bonding pads (on the left) is capacitively coupled to the end of a high-quality co-planar waveguide resonator (on the right). \textbf{b}, View of the design in the area indicated by the red rectangle in \textbf{a}. \textbf{c}, Close view of the \red{3$\times$3-$\mu$m$^2$} copper block forming the normal metal of the QCR. The block is partially overlapping the end of the resonator centre conductor to induce capacitive coupling. Due to shadow evaporation, there is a 20-nm layer of aluminum below most of the copper parts. \textbf{d}, Close view of the bottom NIS junction. The lithographic junction size is 50$\times$70~nm$^2$, giving rise to an effective junction area of roughly 70$\times$70~nm$^2$ due to the 20-nm aluminum layer. See Supplementary Table~\ref{tab_S2} for the parameters of the optimized QCR sample. \textbf{e}, SEM image of a sample similar to the optimized design but with four NIS junctions and a larger normal-metal island.}
\end{figure}

\clearpage

\begin{figure}[p!] \center
	
	\includegraphics[width=1\linewidth]{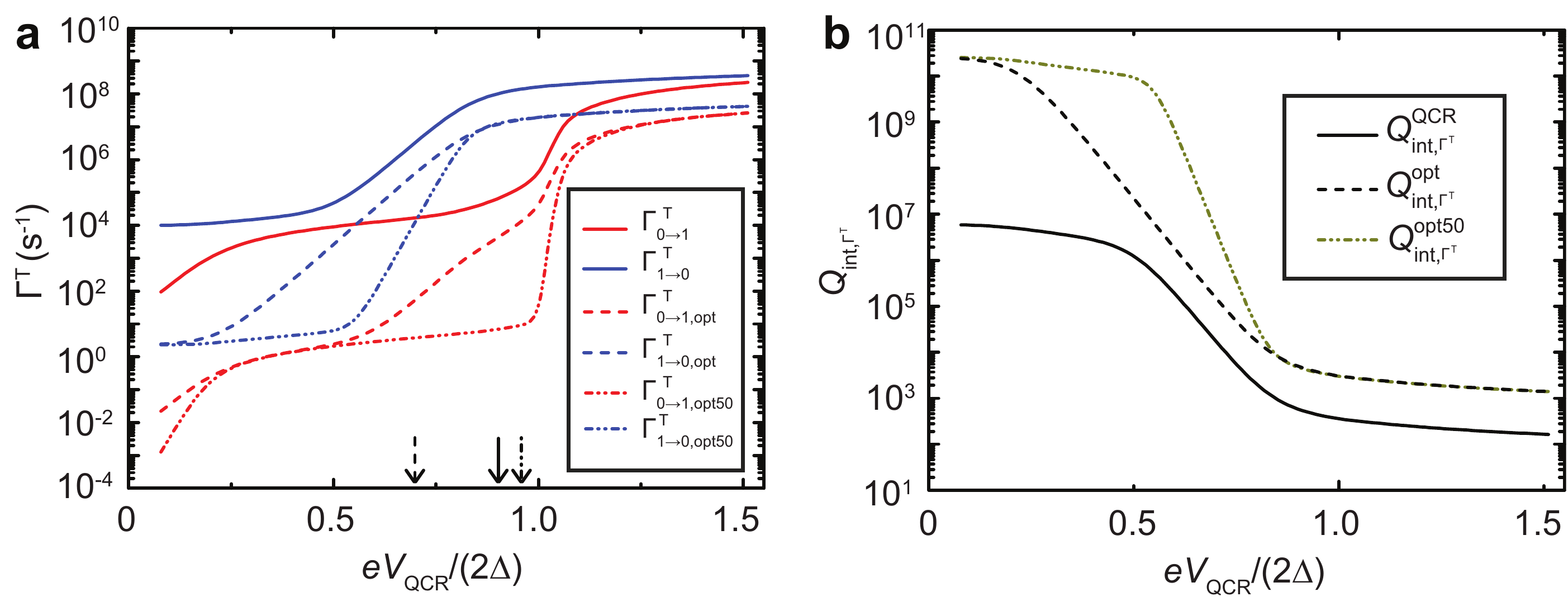}
	\caption{ \label{figS3} \textbf{Effect of operation voltage to the operation characteristics of the QCR.} \textbf{a}, Resonator excitation ($\Gamma^{\mathrm{T}}_{0\rightarrow1}$) and relaxation rates ($\Gamma^{\mathrm{T}}_{1\rightarrow0}$) as functions of the QCR operation voltage, $V_{\mathrm{QCR}}$, for the measured QCR sample (solid lines), the optimized sample using the measured QCR electron temperature (dashed lines), and the optimized sample using 50~mK lower electron temperatures (dash-dotted lines). See Supplemenraty Table~\ref{tab_S2} for the parameters of the optimized sample. \red{Each} operation voltage yielding the minimum temperature corresponding to the photon-assisted tunneling, \red{$T_{\mathrm{res,\Gamma^\mathrm{T}}}=\hbar\omega_0/[\mathrm{log}(\Gamma^{\mathrm{T}}_{0\rightarrow 1}/\Gamma^{\mathrm{T}}_{1\rightarrow 0})k_\mathrm{B}]$, is denoted by} an arrow. Here, the temperature assumes the value 60~mK (solid line), 50~mK (dashed line), and 31~mK (dash-dotted line). \textbf{b}, Resonator quality factor corresponding to the relaxation induced by the photon-assisted tunneling, $Q_{\Gamma^\mathrm{int,T}}$, as a function of the QCR operation voltage for the three cases shown in \textbf{a}. }
\end{figure}

\clearpage

\begin{figure}[p!] \center
	\includegraphics[width=1\linewidth]{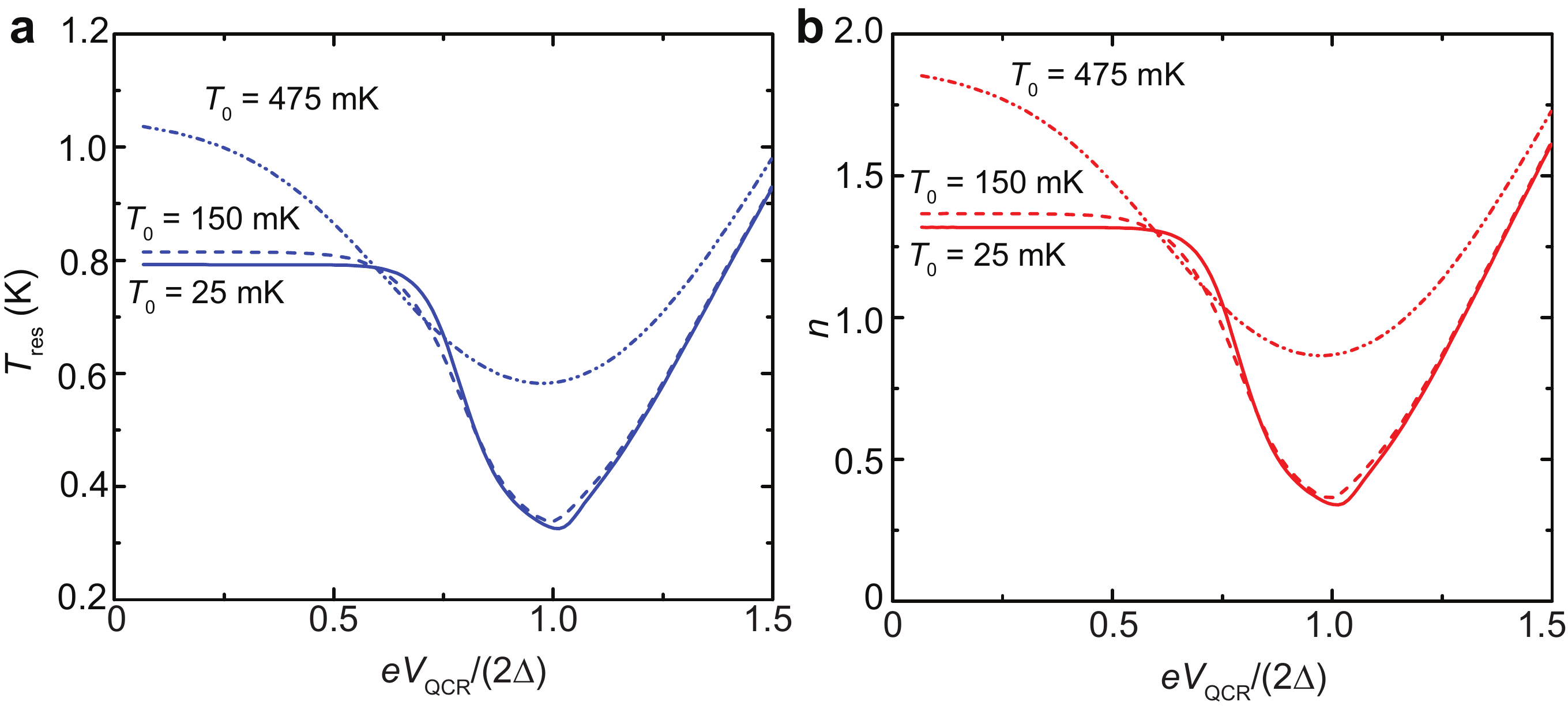}
	\caption{ \label{figS4} \textbf{Resonator temperature and average photon number.} \textbf{a}, Resonator temperature and \textbf{b}, average photon number as functions of the QCR operation voltage at phonon bath temperature of 25 mK (solid lines), 150 mK (dashed lines), and 475 mK (dash-dotted lines). The results are obtained using an upgraded thermal model where the two-state approximation for the resonator mode is not utilized. Here, we employ the experimental data of the QCR electron temperature. The simulation parameters are given in Supplementary Table~\ref{tab_S2} except for $T_\mathrm{x}^\mathrm{sat} = 64$~mK, $\Gamma_\mathrm{leak} = 4.5 \times 10^7$~s$^{-1}$, $\Omega_{\mathrm{QCR}} = 0.01~\mu$m$^3$, and $G_\mathrm{x} = 1.2\times10^{-14}$ W/K. }
\end{figure}
\clearpage

\begin{figure}[p!] \center
	\includegraphics[width=0.5\linewidth]{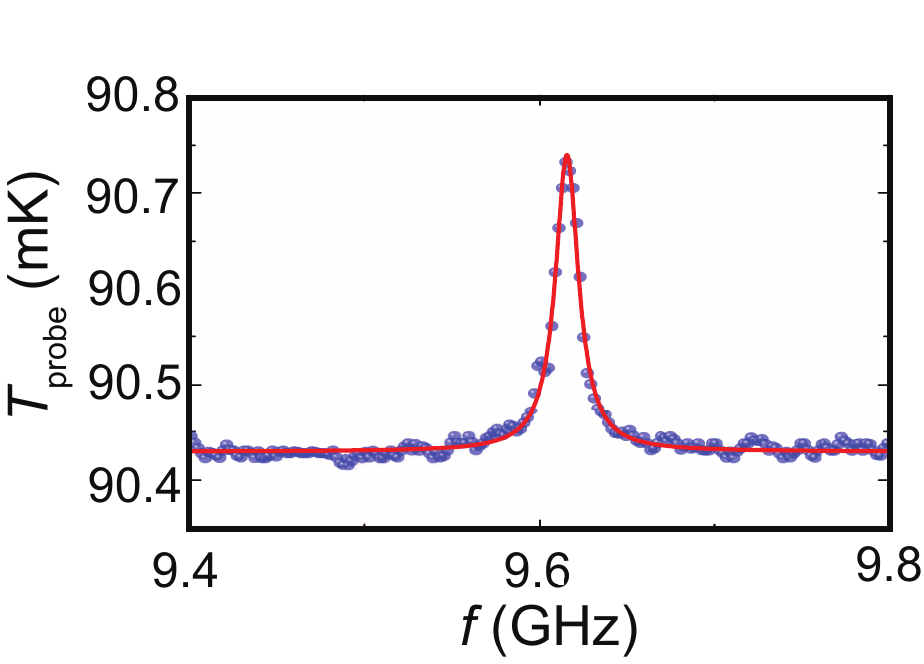}
	\caption{ \label{figS5} \textbf{Quality factor of the control sample.} Experimentally observed electron temperature at the probe resistor (dots) as a function of the frequency of the external microwave excitation. See Fig.~\ref{fig1}a for the measurement scheme. The input power is -56 dBm at room temperature and it is attenuated according to Supplementary Fig.~\ref{figS1}c before reaching the sample. The solid line is a Lorenzian fit to the data. }
\end{figure}

\clearpage

\begin{table} [p!]{
		\centering
		\caption[Table caption text]{\textbf{Device and model parameters.} Identical parameters are used for the active sample and for the control sample unless the specific parameter value for the control sample is given in parenthesis.  \label{tab_S1}}
		\begin{tabular}{ l | c | c | c }
			\hline
			Parameter & Symbol & Value & Unit \\
			\hline \hline
			Resonator length& $L$  & 6.833 & mm \\
			Inductance per unit length & $L_\mathrm{l}$  & $4.7\times 10^{-7}$ &  H/m \\
			Capacitance per uni length & $C_\mathrm{l}$  & $1.3\times 10^{-10}$  &  F/m \\
			Fundamental resonance frequency& $f_0$  & 9.32 & GHz  \\
			Resistance of QCR and probe resistors & $R$  & $46$ & $\Omega$ \\
			Distance of the resistors from resonator edge& $x$  & $100$ & $\mu$m \\
			Volume of QCR and probe resistors &  $\Omega_\mathrm{QCR},\Omega_\mathrm{probe}$ & $4200\times250\times20$ & nm$^3$ \\
			Superconductor gap parameter & $\Delta$  & 214 (216) &  $\mu$eV\\
			Dynes parameter & $\gamma_\mathrm{D}$  & $1\times 10^{-4}$ &  \\
			Normal state junction resistance & $R_\mathrm{T}$  & 23.4 (20.5) & k$\Omega$ \\
			Thermometer bias current & $I_\mathrm{th, QCR}$, $I_\mathrm{th, probe}$  & 17 & pA  \\		
			Material parameter for Cu & $\Sigma_{\mathrm{Cu}}$  & 2~$\times$~10$^{9}$ & W~K$^{-5}$~m$^{-3}$ \\
			Residual heating constant &$\alpha$ & $1.5\times10^{-3}$ &  K$^{-1}$ \\
			Residual heating constant &$\beta$  & 0.38 &  K \\
			Resonator constant excitation rate& $\Gamma_{\mathrm{leak}}$  & 8.062~$\times$~10$^7$ & s$^{-1}$ \\
			Heat conductance to excess bath& $G_{\mathrm{x}}$  & 8.8695$\times$10$^{-14}$ & WK$^{-1}$ \\
			Excess bath saturation temperature& $T_{\mathrm{x}}^\mathrm{sat}$  & 65.4 (104.5) & mK \\
			\hline
		\end{tabular}
		
	}
\end{table}

\clearpage
\begin{table} [p!]{
		\centering
		\caption[Table caption text]{\textbf{Parameters for the optimized QCR sample.} The resistance of the optimized sample (see Supplementary Fig.~\ref{figS2}) is estimated using the measured resistivity of copper in the realized sample and the revised dimensions of the copper block (3000$\times$3000$\times$200~nm$^3$). The total capacitance of the two NIS junctions in the optimized sample, $C_\mathrm{J}^{\mathrm{tot}}$, is obtained using a conveniently realizable junction area of 70$\times$70~nm$^2$ and the usual junction capacitance per unit area 45 fF/$\mu$m$^2$. The effective distance of the QCR from the edge of the resonator is calculated from the total junction capacitance as described above. The value of the Dynes parameter is obtained from ref.~\citen{SairaPRB2012} for NIS junctions with proper filtering and shielding. The normal-state junction resistance can be increased compared with the realized sample by increasing the oxidation time and pressure in the fabrication process. \label{tab_S2}}
		\begin{tabular}{ l | c | c | c }
			\hline
			Parameter & Symbol & Value & Unit \\
			\hline \hline
			Resonator length& $L$  & 6.833 & mm \\
			Inductance per unit length & $L_\mathrm{l}$  & $4.7\times 10^{-7}$ &  H/m \\
			Capacitance per uni length & $C_\mathrm{l}$  & $1.3\times 10^{-10}$  &  F/m \\
			Fundamental resonance frequency& $f_0$  & 9.32 & GHz  \\
			Resistance of QCR resistor & $R$  & $0.3$ & $\Omega$ \\
			Total junction capacitance &$C_\mathrm{J}^{\mathrm{tot}}$ & 440 &  aF \\
			Distance of the resistors from resonator edge& $x$  & $3.4$ & $\mu$m \\
			Volume of QCR resistor &  $\Omega_\mathrm{QCR}$ & $3000\times3000\times200$ & nm$^3$ \\
			Superconductor gap parameter & $\Delta$  & 214 &  $\mu$eV\\
			Dynes parameter & $\gamma_\mathrm{D}$  & $2\times 10^{-7}$ &  \\
			Normal state junction resistance & $R_\mathrm{T}$  & 200 & k$\Omega$ \\		
			Material parameter for Cu & $\Sigma_{\mathrm{Cu}}$  & 2~$\times$~10$^{9}$ & W~K$^{-5}$~m$^{-3}$ \\
			\hline
		\end{tabular}
		
	}
\end{table}

\clearpage

\vspace{10 pt}\noindent{\bf Supplementary References}\label{Supp:PE}

\clearpage

\end{document}